\documentclass[groupedaddress,aps,prb,twocolumn,showpacs,amsmath,amssymb]{revtex4-1}

\usepackage{graphicx}% Include figure files
\usepackage{dcolumn}% Align table columns on decimal point
\usepackage{bm}% bold math
\usepackage[normalem]{ulem} %This is for "strike through" capability
\usepackage [utf8] {inputenc}
\usepackage{natbib}

\newcommand{\ket}[1]{\left|\kern +0.2ex  #1 \right>} % for Dirac bras
\newcommand{\bra}[1]{\left< #1 \kern +0.2ex  \right|} % for Dirac kets
\newcommand{\braket}[2]{\left<\kern +0.2ex #1\kern +0.2ex  |\kern +0.2ex  #2 \kern +0.2ex\right>} % for Dirac kets

\def\vec#1{\textbf{#1}}

\newcommand{\YI}[1]{Y_{\vec{#1},\tau}^{\rm I}}
\newcommand{\YII}[1]{Y_{\vec{#1},\tau}^{\rm II}}
\newcommand{\I} {{\rm I}}
\newcommand{\II} {{\rm II}}

\newcommand{\kpar}{\vec{k}_\|}
\newcommand{\csub}{c_{\rm sub}}
\newcommand{\csup}{c_{\rm sup}}
\newcommand{\kz}[1]{k_z^{\vec{n}}}
\newcommand{\kzI}[1]{k_z^{\vec{n},\I}}
\newcommand{\kzII}[1]{k_z^{\vec{n},\II}}

\begin{document}

\title{Optical properties of Fano-resonant metallic metasurfaces on a substrate}
\author{S. Hossein Mousavi}
%\affiliation{Department of Physics and Institute for Fusion Studies, University of Texas at Austin, One University Station C1500, Austin, Texas 78712, USA}
\author{Alexander B. Khanikaev}
%\affiliation{Department of Physics and Institute for Fusion Studies, University of Texas at Austin, One University Station C1500, Austin, Texas 78712, USA}
\author{Gennady Shvets}
\email{gena@physics.utexas.edu}               
\affiliation{Department of Physics and Institute for Fusion Studies, University of Texas at Austin, One University Station C1500, Austin, Texas 78712, USA}
\begin{abstract}
Three different periodic optical metasurfaces exhibiting Fano resonances are studied in mid-IR frequency range in the presence of a substrate. We develop a rigorous semi-analytical technique and calculate how the presence of a substrate affects optical properties of these structures. An analytical minimal model based on the truncated exact technique is introduced and is shown to provide a simple description of the observed behavior. We demonstrate that the presence of a substrate substantially alters the collective response of the structures suppressing Wood's anomalies and spatial dispersion of the resonances. Different types of Fano resonances are found to be affected differently by the optical contrast between the substrate and the superstrate. The dependence of the spectral position of the resonances on the substrate/superstrate permittivities is studied and the validity of the widely used effective medium approaches is re-examined. 
\end{abstract}

\maketitle  
\section{Introduction} 
Optical nanostructures exhibiting Fano resonances have attracted a significant attention of the research community in recent years~\cite{miroshnichenko_fano_2010, lukyanchuk_fano_2010, wu_broadband_2011, fedotov_sharp_2007, papasimakis_coherent_2009, zhang_plasmon-induced_2008, liu_plasmonic_2009, rybin_fano_2009, verellen_fano_2009, fan_self-assembled_2010, fan_fano-like_2010, hao_tunability_2009}. As classical analogues of quantum Fano system~\cite{fano_effects_1961}, these systems represent elegant tabletop tools for testing fundamental principles of physics~\cite{alonso-gonzalez_real-space_2011}. From the applications' point of view, the possibility to trap, enhance, and manipulate light in optical Fano nanostructures and metamaterials is also very promising and has already been demonstrated to be beneficial in sensing and bio-sensing~\cite{lahiri_asymmetric_2009,adato_ultra-sensitive_2009,yanik_seeing_2011,wu_fano-resonant_2011}, photovoltaics and thermo-photovoltaics~\cite{atwater_plasmonics_2010,rephaeli_absorber_2009, ghebrebrhan_tailoring_2011}, and slow-light generation~\cite{wu_broadband_2011}. 

\begin{figure}[htbp]
	\centering
	\includegraphics[draft=false,width=.48\textwidth]{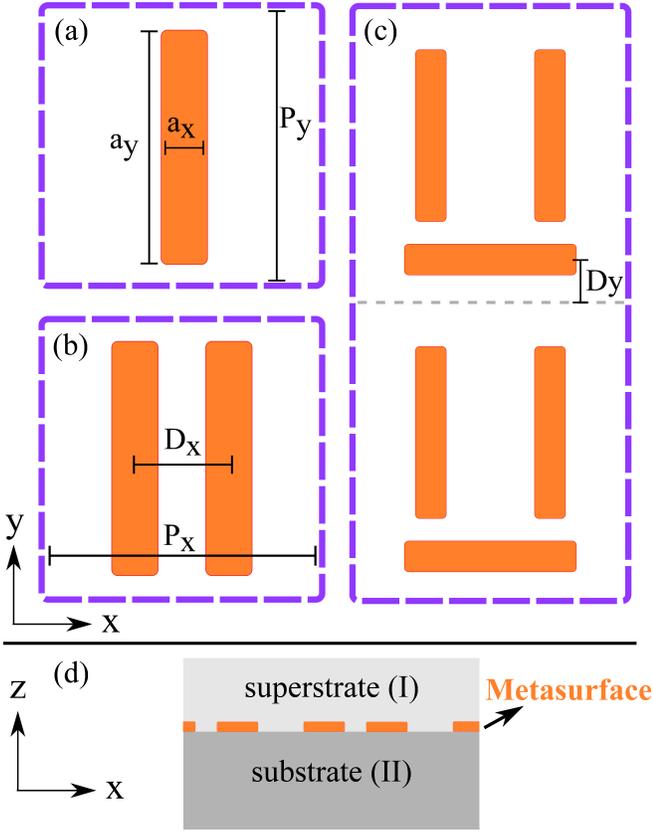}
	\caption{(Color online.) Schematic unit cell of the three structures considered in this paper: (a) single-antenna, (b) double-antenna, and (c) dolmen metasurfaces. In (c), two unit cells (separated by a dashed line) are shown. (d) shows the side view of the metasurfaces cladded by the superstrate and the substrate.}
	\label{Fig:Schematics}	
\end{figure}	

Fano resonances in optical metamaterials originate from electromagnetic interactions between their constituents. According to the type of interaction, Fano resonances can be classified into two groups: (a) coherent Fano resonances which are the result of interferences between all of the meta-atoms forming a periodic array, and (b) local Fano resonances which originate from the complex local structure of individual meta-atoms. While for the first class of these resonances, periodic arrangement of a large number of meta-molecules is crucial, for the second class, even a single meta-molecule may exhibit a Fano resonance~\cite{lukyanchuk_fano_2010}. However, in both cases the response of the meta-surface will depend on the dielectric environment. For coherent metamaterials, it is well-known that the presence of a substrate modifies the far-field interactions\cite{ford_electromagnetic_1984, aizpurua_substrate-enhanced_2008}, which substantially alters the collective response. However, even in the case of local Fano resonances, the presence of a substrate can significantly modify the interaction between the meta-molecule's constituents\cite{powell_substrate-induced_2010}. Thus, understanding of substrate effects in both cases is critical for basic understanding of Fano resonances in experimentally relevant systems.

 The geometry of the systems with Fano resonances can vary from simple designs such as metal nanoparticles~\cite{lukyanchuk_fano_2010} and perforated films~\cite{fano_JOSA_1941, genet_fano-type_2003,sarrazin_role_2003} to structures with complex geometries such as oligomers and dolmens~\cite{verellen_fano_2009,fan_self-assembled_2010,fan_fano-like_2010,alonso-gonzalez_real-space_2011}.
 
 The simplest approach to describe the Fano resonances is a mechanical toy model of coupled harmonic oscillators ~\cite{zhang_plasmon-induced_2008, liu_plasmonic_2009, petschulat_simple_2010, wu_broadband_2011}. In many cases, such model provides sufficient insight into the physics of the system; it allows one to fit optical spectra and associate spectral features of the system with the known properties of the mechanical Fano model. Nevertheless, such simplified representation of a complex physical system neglects various effects relevant for real structures. Among these are substrate effects, effects of periodicity resulting in a collective character of the electromagnetic response of meta-materials, geometry/proximity effects modifying the interactions between meta-molecules. 
 
%Despite the complexity of the systems, one of the most common approaches to describe the Fano resonances is to use the simple mechanical model of coupled harmonic oscillators ~\cite{zhang_plasmon-induced_2008, liu_plasmonic_2009, petschulat_simple_2010, wu_broadband_2011}. In many cases such model provides a sufficient insight into the physics of the system; it allows one to fit optical spectra and associate the spectral features of the system with the known properties of the mechanical Fano model. While being useful, such simplified representation of the complex physical system inevitably results in a loss of the physical ground and impossibility to consider various effects relevant for real structures. Among them are substrate effect, effects of periodicity resulting in a collective character of the electromagnetic response of meta-materials, geometry and proximity effects modifying the character of interaction between meta-molecules, etc. 

Another analytical description applicable to some relatively simple systems, such as arrays of point-like dipoles, is offered by a dipole model~\cite{zou_narrow_2004, zou_silver_2004, auguie_collective_2008, garcia_de_abajo_colloquium:_2007,papasimakis_coherent_2009}. Indeed, a few of the above-mentioned phenomena, including collective effects and Fano resonances, can be adequately described by this model~\cite{mousavi_suppression_2011}. However, there are situations when the dipole model is inadequate. For instance, as soon as the scatterers or the metamolecules forming the array become comparable in size to the wavelength of light, the dipole-dipole interaction mechanism breaks down because interactions through higher multipoles come into play. Incorporating substrate effects also represents a challenge for the dipole model. To resolve these problems, the dipole model has been extended by including effects of higher multipoles~\cite{stafanou_scattering_1991}, found with the use of numerically calculated polarizabilities of metamolecules and by applying the scattering-matrix technique~\cite{stafanou_scattering_1992, bendaa_confined_2009, auguie_diffractive_2010}. However, after such generalizations, the dipole model can no longer be considered analytical.
To fully account for all the above-mentioned effects, the approach which is being widely used is to apply a powerful yet time consuming ab-initio numerical solvers of the Maxwell equations, or various generalizations of hybrid numerical/analytical approaches~\cite{gallinet_ab_2011}.

However, there is another very promising semi-analytical yet rigorous approach based on the modal-matching technique (MMT). MMT has been proven to be very fruitful for describing various electromagnetic systems ranging from frequency-selective surfaces and antenna arrays in radio frequency domain~\cite{dawes_thin_1989} to perforated metallic structures~\cite{martin-moreno_theory_2001, hendry_importance_2008, mousavi_highly_2010,khanikaev_one-way_2010}. In both cases, the approach is relying on the expansion of a system-specific polarization (e.g., the total current in the antennas or the fields inside the holes) in the form of a superposition of the eigenmodes satisfying appropriate boundary conditions. It is important that, because the basis used for the expansion takes into account meta-molecules' geometry, this approach can account for the geometry-specific effects, including shape resonances and proximity effects~\cite{powell_metamaterial_2010,giessen_opex_2011}. %Note that in the case of the perforated metal films the plasmonic effects have also been successfully taken into account in modal matching approach by applying the impedance boundary condition approximation [18]. 

In this paper, the semi-analytical MMT based on the current expansion~\cite{dawes_thin_1989} is applied to study Fano resonances in two-dimensional periodic arrays of metallic antennas on a substrate (Fig.~\ref{Fig:Schematics}). The emphasis is on arrays of complex antennas resonant in the mid-IR part of optical spectrum, which is important for bio-sensing~\cite{adato_ultra-sensitive_2009,wu_fano-resonant_2011} and thermo-photovoltaic applications~\cite{atwater_plasmonics_2010,rephaeli_absorber_2009, ghebrebrhan_tailoring_2011}. MMT is shown to capture the meta-molecule geometry as well as the substrate effects and is easily expandable to the cases of more complex geometries composed of several spatially extended scatterers per unit cell. We develop a minimal model of the MMT by truncating the electric-current basis of the full model and demonstrate that even the resulting analytical model is capable to quantitatively describe the optical properties of the structures. The main emphasis of the paper is on the effect of the substrate. It is demonstrated that with the model in hand the MMT can predict the spectral positions of the resonances and describe the interference among the different scattering pathways provided by the different metamolecule's resonances.

The rest of the paper is organized as follows. In Section II a general MMT based on the Rayleigh and current expansions is derived. In Section III a minimal model is developed and applied to study a periodic single-antenna metasurface (SAM), shown in Fig.~\ref{Fig:Schematics}a. %This structure is complementary counterpart of the optically thin perforated metal film exhibiting an extraordinary optical transmission, and represents the simplest periodic Fano-resonant metasurface. Then we study more complex Fano systems. 
In Section IV we analyze periodic double-antenna meta-surfaces (DAMs) [Fig.~\ref{Fig:Schematics}b], which are known to exhibit a Fano resonance due to the interference of the modes corresponding to symmetric and anti-symmetric charge distributions in the antenna pairs. Finally, in Section V a dolmen structure formed by three antennas (Fig.~\ref{Fig:Schematics}c) and known to exhibit a plasmonic analogue of electromagnetically induced transparency (EIT)~\cite{zhang_plasmon-induced_2008,liu_plasmonic_2009} is studied.

\section{Modal Matching Technique: Theoretical Formalism}
In this section the electric current expansion technique is generalized to the case of infinitesimally thin metallic antennas of finite surface conductivity. Following Ref.~\cite{dawes_thin_1989}, we rely on the current expansion, but require the current in the plasmonic antennas to be defined by the high-frequency conductivity of the metal and the electric field right on the antennas' surface. To make the expressions more compact and keep the formulation more general, the Dirac notation is used.
Using the periodicity of the structures the tangential components of the fields are expanded in the superstrate ($\I$) and the substrate ($\II$) in the plane-wave basis,
	\begin{align}
		\vec{E}_\|^\I&=\sum_{\vec{n},\tau}{(i_{\vec{n},\tau} e^{i \kzI{n}z}+r_{\vec{n},\tau} e^{-i \kzI{n}z})\ket{\vec{n},\tau}},\notag\\
		\vec{E}_\|^\II&=\sum_{\vec{n},\tau}{t_{\vec{n},\tau} e^{i \kzII{n} z}\ket{\vec{n},\tau}}, \notag\\
		-\hat{z}\times\vec{H}_\|^\I  &= \sum_{\vec{n},\tau}{\YI{n} (i_{\vec{n},\tau} e^{i \kzI{n} z}-r_{\vec{n},\tau} e^{-i \kzI{n} z})\ket{\vec{n},\tau}},\label{eqn:fields}\\
		-\hat{z}\times\vec{H}_\|^\II &= \sum_{\vec{n},\tau}{\YII{n} t_{\vec{n},\tau} e^{i \kzII{n} z}\ket{\vec{n},\tau}},\notag
	\end{align}
where $\ket{\vec{n},\tau}=\ket{n_x,n_y,\tau}$ represents the in-plane (tangential) electric field of the $\vec{n}$-th diffracted plane wave with an in-plane wave-number $\vec{k}_{\|}^{\vec{n}}=(k_{x}^{\vec{n}},k_{y}^{\vec{n}})$, a polarization state $\tau$ ($s$ or $p$ polarization) and a wave admittance in the superstrate (substrate) $\YI{n}$ ($\YII{n}$). The wave admittance of an $s$-polarized plane wave is $Y_{\vec{n},s}=\kz{n}/(Z_0k_0)$ and that of a $p$-polarized plane wave is $Y_{\vec{n},p}=\epsilon_r k_0/(Z_0\kz{n})$. In the above expressions, $k_x^\vec{n}=2\pi n_x/P_x$, $k_y^\vec{n}=2\pi n_y/P_y$, and $k_z^\vec{n}=\sqrt{\epsilon_r k_0^2-(k_\|^\vec{n})^2}$ are respectively the $x$, $y$, and $z$ components of the wavevector $\vec{k}$. $k_0$ is the wavenumber of the wave in vacuum and $\epsilon_r$ is the relative permittivity of the medium in which the wave is propagating. $Z_0\approx376.73~\Omega$ is the impedance of vacuum. The spatial representation of $\ket{\vec{n},\tau}$ is given by:
	\begin{align}
	\braket{\vec{r}_\|}{\vec{n},s}=\frac{\exp{(ik_x^\vec{n}x+ik_y^\vec{n}y)}}{k_\|^\vec{n}\sqrt{P_xP_y}} \{k_y^\vec{n}\hat{x}-k_x^\vec{n}\hat{y}\}, \\
  \braket{\vec{r}_\|}{\vec{n},p}=\frac{\exp{(ik_x^\vec{n}x+ik_y^\vec{n}y)}}{k_\|^\vec{n}\sqrt{P_xP_y}} \{k_x^\vec{n}\hat{x}+k_y^\vec{n}\hat{y}\}.
	\end{align}
The surface current in the plasmonic antennas is expanded with the use of the basis functions $\ket{\vec{j}_\alpha^m}$ satisfying the boundary condition of vanishing current on the antenna edges. The electric current in the $m$-th antenna within the unit cell takes the form 
	\begin{equation}
		\vec{J}^m=\sum_\alpha{c_\alpha^m \ket{\vec{j}_\alpha^m}}
	\label{eqn:jm}
	\end{equation}
where $c_\alpha^m$ is the amplitude of the $\alpha$-th current mode. The precise form of the functions $\ket{\vec{j}_\alpha^m}$ depends on the antenna geometry, which will be taken rectangular throughout the paper. For this case, the explicit expressions for $\ket{\vec{j}_\alpha^m}$ are given elsewhere.~\cite{dawes_thin_1989}

The finite conductivity of the antennas is incorporated into the model using the relation
	\begin{equation}
  	\vec{J}^m=\sigma\vec{E}_\|^m
   \label{Eq:JsE}
    \end{equation}
where $\sigma=ih \epsilon_m k_0$ is the effective surface conductivity of the antennas of thickness $h$ and permittivity $\epsilon_m$~\cite{vakil_transformation_2011}. 
The fields in the substrate and the superstrate should satisfy the standard continuity boundary condition over the interface between them
	\begin{subequations}
	\begin{align}
		\vec{E}_\|^\I               &=\vec{E}_\|^\II,                \label{Eq:EOverDiel}\\ 
        -\hat{z}\times\vec{H}_\|^\I &=-\hat{z}\times\vec{H}_\|^\II.  \label{Eq:HOverDiel}
    \end{align}
    \label{Eq:Fields}
Over the antennas, the electric field is continuous and Eq.~(\ref{Eq:EOverDiel}) still holds, however a discontinuity in the magnetic field appears due to the finite current in the antennas
	\begin{align} 
	-\hat{z}\times\left(\vec{H}_\|^\I-\vec{H}_\|^\II\right)=\vec{J}. \label{Eq:HOverMetal}
	\end{align}
	\end{subequations}
Combining these constraints together with the modal expansions and using the orthonormality of the basis functions, we obtain a system of three linear equations:
	\begin{subequations}
	\begin{align}
		&i_{\vec{n},\tau}+r_{\vec{n},\tau}=t_{\vec{n},\tau},\\
		&\YI{n} (i_{\vec{n},\tau}-r_{\vec{n},\tau} )-\YII{n} t_{\vec{n},\tau}=\sum_{\alpha,m}{\braket{\vec{n},\tau}{{\vec{j}_\alpha^m}} c_\alpha^m},\\
		&\sum_{\vec{n},\tau}{\braket{\vec{j}_\alpha^m}{\vec{n},\tau}(i_{\vec{n},\tau}+r_{\vec{n},\tau})}=\sigma^{-1} c_\alpha^m.
	\end{align}
	\label{Eq:system}
	\end{subequations}
Eqs.~(\ref{Eq:system}a-\ref{Eq:system}b) are obtained by multiplying Eqs.~(\ref{Eq:Fields}a-c) (from left) by $\bra{\vec{n},\tau}$ and Eq.~(\ref{Eq:system}c) is obtained by multiplying Eq.~(\ref{Eq:JsE}) (from left) by $\bra{\vec{j}_\alpha^m}$. The scalar product $\braket{\vec{n},\tau}{\vec{j}}$ is defined as
\begin{align}
\braket{\vec{n},s}{\vec{j}}=&\int_{-a_x/2}^{a_x/2}\int_{-a_y/2}^{a_y/2}\frac{\exp{(-ik_x^\vec{n}x-ik_y^\vec{n}y)}}{k_\|^n\sqrt{P_xP_y}}\times\\
&\{k_y^\vec{n}j_x(x,y)-k_x^\vec{n}j_y(x,y)\}dx\,dy, \notag\\
\braket{\vec{n},p}{\vec{j}}=&\int_{-a_x/2}^{a_x/2}\int_{-a_y/2}^{a_y/2}\frac{\exp{(-ik_x^\vec{n}x-ik_y^\vec{n}y)}}{k_\|^n\sqrt{P_xP_y}}\times\\
&\{k_x^\vec{n}j_x(x,y)+k_y^\vec{n}j_y(x,y)\}dx\,dy.\notag
\end{align}
where $j_x(x,y)$ and $j_y(x,y)$ are the $x$ and $y$ components of the current profile $\ket{\vec{j}}$. Eqs.~(\ref{Eq:system}a-\ref{Eq:system}c) can be solved to obtain the amplitude of the scattered fields $t_{\vec{n},\tau}$ and $r_{\vec{n},\tau}$ in terms of the antenna current modes $c_\alpha^m$
	\begin{align}
	t_{\vec{n},\tau}=\frac{2\YI{n}}{\YI{n}+\YII{n}}i_{\vec{n},\tau}-\frac{\sum_{\alpha,m}{\braket{\vec{n},\tau}{{\vec{j}_\alpha^m}} c_\alpha^m}}{{\YI{n}+\YII{n}}},\label{Eq:tn}\\
	r_{\vec{n},\tau}=\frac{\YI{n}-\YII{n}}{\YI{n}+\YII{n}}i_{\vec{n},\tau}-\frac{\sum_{\alpha,m}{\braket{\vec{n},\tau}{{\vec{j}_\alpha^m}} c_\alpha^m}}{{\YI{n}+\YII{n}}},\label{Eq:rn}
	\end{align}
where the amplitude of the $\alpha$-th current mode in the $m$-th antenna $c_\alpha^m$ is
	\begin{align}
		\sum_{\alpha\prime,m\prime} \lbrace\sigma^{-1}&\delta_{\alpha,\alpha\prime}\delta_{m,m\prime}+S_{\alpha}^{m}{}_{\alpha\prime}^{m\prime}\rbrace c_{\alpha\prime}^{m\prime}=\chi_{\alpha}^{m}\label{Eq:c}
	\end{align}
In Eq.~(\ref{Eq:c}), $S_{\alpha}^{m}{}_{\alpha\prime}^{m\prime}$ is the Green's function that describes the cross-talk between the ${\alpha}$-th current mode of the $m$-th antenna with the ${\alpha\prime}$-th current mode of the $m\prime$-th antenna in the unit cell. This coupling is mediated by the plane waves of all (propagating and evanescent) diffraction orders $\vec{n}$ as given by Eq.~(\ref{Eq:S}). $\chi_{\alpha}^m$ is the direct coupling strength of the ${\alpha}$-th current mode of the $m$-th antenna to the external field.
\begin{subequations}
	\begin{align}
		S_{\alpha}^{m}{}_{\alpha\prime}^{m\prime}&=\sum_{\vec{n},\tau}{\frac{\braket{\vec{j}_\alpha^m}{\vec{n},\tau}\braket{\vec{n},\tau}{\vec{j}_{\alpha\prime}^{m\prime}}}{\YI{n}+\YII{n}}}\label{Eq:S}\\
		\chi_{\alpha}^{m}&=\sum_{\vec{n},\tau}{\frac{2\YI{n}\braket{\vec{j}_\alpha^m}{\vec{n},\tau}}{\YI{n}+\YII{n}}i_{\vec{n},\tau}}.\label{Eq:chi}
		\end{align}
	\end{subequations}
In addition to the scattering characteristics, the eigenmodes of the structure can also be determined by solving the secular equation
	\begin{align}
{\rm det} \lbrace\sigma^{-1}\delta_{\alpha,\alpha\prime}\delta_{m,m\prime}+S_{\alpha}^{m}{}_{\alpha\prime}^{m\prime}\rbrace=0. \label{Eq:DetS}
	\end{align}
In general, Eq.~(\ref{Eq:DetS}) can be satisfied only at complex frequencies $\omega=\omega_r+i\omega_i$, showing that the eigenmodes have a finite lifetime due to either Ohmic losses (when $\Re e\{\sigma\}\neq0$) or radiative decay\cite{mousavi_highly_2010,khanikaev_one-way_2010}. A real eigenvalue $\omega$ may be ideally achieved only in the limit of lossless meta-surfaces and for $k_\parallel>k$, which ensures absence of absorption and radiation. While in this paper we assume a two-dimensional periodic array of the antennas, but the results can be generalized to the case of an individual antenna. In that case, the sum over the discrete set of plane waves $\sum_\vec{n}$ in the above expressions, should be replaced by an integral over the continuum $\int \int dk_x dk_y$.\cite{garcia-vidal_transmission_2006}

\section{Collective response of Single-Antenna Metasurfaces}
While Eqs.~(\ref{Eq:tn}-\ref{Eq:c}) fully characterize the optical properties of the system, they have to be solved numerically. However, in the case when the wavelength of the incident light matches the resonant wavelength of a particular current eigenmode $\ket{\vec{j}_\alpha^m}$, truncation of the current modes' basis can provide simple and instructive expressions~\cite{martin-moreno_minimal_2008}. Indeed, it has been shown for the case of perforated metal films that near the cut-off frequency of a particular waveguide mode such a ``minimal'' model can provide a very good approximation~\cite{garcia_de_abajo_electromagnetic_2005, hendry_importance_2008}. As will be shown later this situation holds for the antenna geometries in the frequency range studied here. It is worthwhile mentioning here, however, that as dimensions of antennas decrease and they become increasingly subwavelength, the convergence of the MMT deteriorates and more and more current modes should be considered.
\begin{figure}[htbp]
	\centering
	\includegraphics[draft=false,width=.48\textwidth]{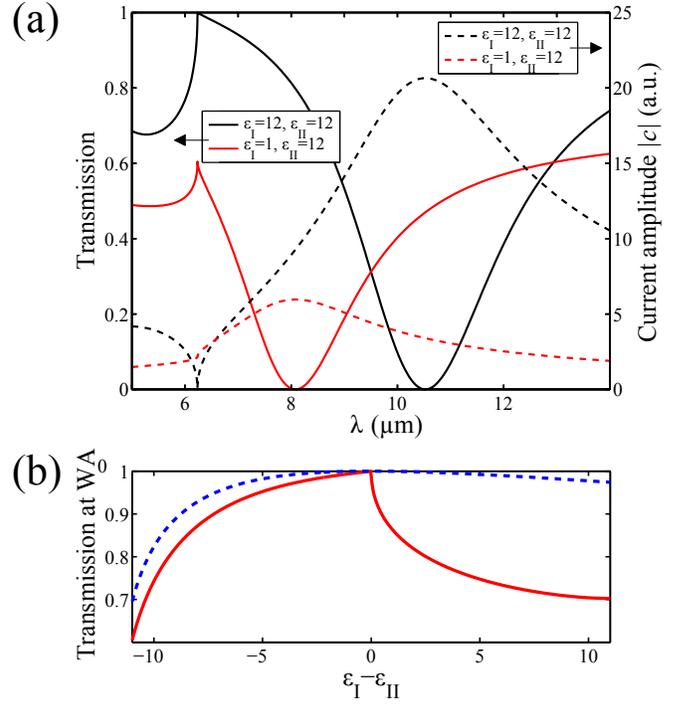}
	\caption{(Color online.) (a) Zeroth-order normal-incidence transmission (solid lines) and amplitudes of the fundamental current mode (dashed lines) for the single-antenna metasurface for the symmetric ($\epsilon_{\rm I}=\epsilon_{\rm II}$, black lines) and asymmetric ($\epsilon_{\rm I}\neq\epsilon_{\rm II}$, red lines) claddings. (b) Zeroth-order transmission of the single-antenna metasurface at the Wood's anomaly as a function of the dielectric contrast $\epsilon_{\rm I}-\epsilon_{\rm II}$(solid line) and the bare (without antennas) interface transmission (dashed line). The structure parameters are as follows: $a_x=0.3~{\mu}m$, $a_y=1.5~{\mu}m$, $P_x=1.8~{\mu}m$, $P_y=1.8~{\mu}m$, and $\epsilon_\II=12$.}
	\label{Fig:SAM}
\end{figure}		

By limiting the current basis to one mode per antenna $\ket{\vec{j}_1^m}=\sqrt{\frac{2}{a_y}}\cos(\frac{\pi y}{a_y})\,\hat{y}$ we assume a dominant role for the fundamental (dipolar) antenna-mode. Such a truncation allows us to get simple analytical expressions for the transmission and reflection amplitudes which qualitatively explain the system's response yet quantitatively match the numerical results.

First we consider a structure with a single antenna per unit cell with large length-to-width aspect ratio so that the fundamental current mode $\ket{\vec{j}}\equiv\ket{\vec{j}_1^1}$ with the current along the long antenna dimension dominates. In this case Eq.~(\ref{Eq:c}) can be analytically solved for the current amplitude
\begin{equation}
c(\kpar,\omega)\equiv c_1^1(\kpar,\omega)=\frac{\chi_1(\kpar,\omega)\,E_{\rm ext}}{\sigma^{-1}(\omega)+S_{11}(\kpar,\omega)},\label{Eq:c1}
\end{equation}
where $\chi_1 (\kpar,\omega)=2 \YI{0}/(\YI{0}+\YII{0})\braket{\vec{j}}{\vec{0},\tau}$ is the coupling efficiency of the antenna to the external field, the sum $S_{11}(\kpar,\omega)\equiv S_{11}^{11}(\kpar,\omega)$ as defined in Eq.~(\ref{Eq:S}), $E_{\rm ext}=i_{\vec{0},\tau}$ and the rest of the $i_{\vec{n},\tau}$'s are assumed to be zero. In section III and IV we assume that the incident polarization is along the long dimension of the antennas. In the minimal model, $S_{11}$ is explicitly given by:
\begin{align}
S_{11}=\frac{4 Z_0a_x^2 a_y}{\pi^2 P_xP_y}\sum_\vec{n}{\frac{{\rm sinc}^2(a_xk_x^\vec{n}/2)(1+\cos(a_yk_y^\vec{n}))}{(1-(a_yk_y^\vec{n}/\pi)^2)^2}}\times\notag\\
\left[\frac{k_0\left(k_x^\vec{n}/k_\|^\vec{n}\right)^2}{\kzI{n}+\kzII{n}}+\frac{\left(k_y^\vec{n}/k_\|^\vec{n}\right)^2}{\epsilon_\I k0/\kzI{n}+\epsilon_\II k0/\kzII{n}}\right],
\end{align}
where the first term in the square bracket corresponds to the $s$-polarized diffraction orders and the second term corresponds to the $p$-polarized diffraction orders. ${\rm sinc}(t)\equiv\sin(t)/t$ can be approximated by unity for thin antennas ($t=k_xa_x/2\approx0$). We emphasize that this sum converges much faster compared to that in the dipole model. In the framework of the minimal model the transmission and reflection coefficients of the single-antenna meta-surface assume a simple form:
\begin{align}
t_{\vec{n},\tau}=\frac{2\YI{n}}{\YI{n}+\YII{n}}i_{\vec{n},\tau}-\frac{\braket{\vec{n},\tau}{\vec{j}}}{\YI{n}+\YII{n}}c,\label{Eq:t1}\\
r_{\vec{n},\tau}=\frac{\YI{n}-\YII{n}}{\YI{n}+\YII{n}}i_{\vec{n},\tau}-\frac{\braket{\vec{n},\tau}{\vec{j}}}{\YI{n}+\YII{n}} c.\label{Eq:r1}
\end{align}
Eqs.~(\ref{Eq:t1}, \ref{Eq:r1}) explicitly show two scattering pathways: (i) the direct transmission (reflection) through the dielectric interface between the substrate and the superstrate without interacting with the antennas and (ii) the field radiated by the antennas due to the excitation of the current mode. These two scattering pathways and the presence of the Wood's anomaly~\cite{Wood, Rayleigh} lead to a Fano interference resulting in an asymmetric lineshape typically observed in antenna arrays (Fig.~\ref{Fig:SAM}a) and systems with extraordinary optical transmission. The condition of the vanishing denominator in the Eq.~(\ref{Eq:c1}), $\sigma^{-1}(\omega)+S_{11}(\kpar,\omega)=0$, approximately describes the eigenmodes dispersion of the antenna array (\textit{cf.} Eq.~(\ref{Eq:DetS}) for the exact expression).
\begin{figure}[htbp]
	\centering
	\includegraphics[draft=false,width=.48\textwidth]{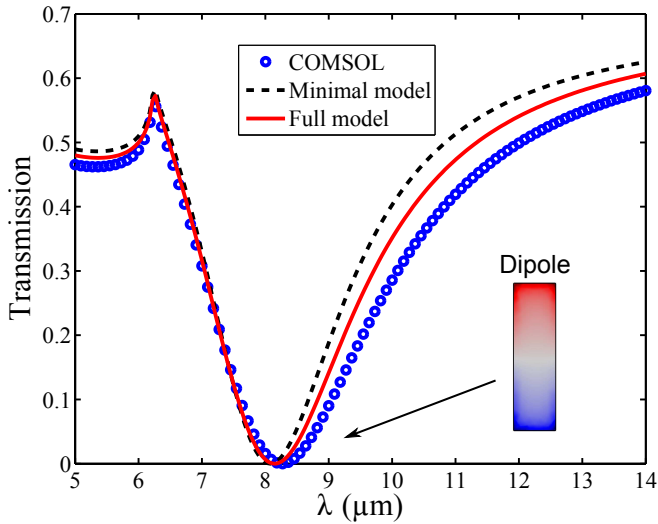}
	\caption{(Color online.) Direct transmission at normal incidence calculated using three methods: (i) first-principles COMSOL simulations (circles), (ii) full MMT model (solid line), and (iii) the minimal MMT model (dashed line). Structure parameters: same as in Fig.~\ref{Fig:SAM}.}
	\label{Fig:ComsolSAM}
\end{figure}

Note that Eq.~(\ref{Eq:c1}) strongly resembles that of the effective polarizability of an array of dipoles in the dipole model~\cite{garcia_de_abajo_colloquium:_2007}. In our case, $\Im m\{1/\sigma(\omega)\}$ plays the role of ``plasmonic polarizability'' of the antennas while the role of $S_{11}(\kpar,\omega)$ is similar to that of the dipole lattice sum. Note that because the modal-matching technique is based on current expansion, the polarizability and the lattice sum have $\pi/2$ phase shift as compared to the dipole model. It can be shown that when the dipole sum of the dipole model is written in the reciprocal space, the sums of both models have similar terms diverging at the Wood's anomalies. However, in contrast to the dipole model, the antenna model fully considers the substrate, antenna shape and finite conductivity. Here we limit our consideration to the mid-IR domain where antennas can be considered as perfectly conducting ($1/\sigma$=0). Effect of the finite conductivity of the antennas on Fano resonances will be considered elsewhere.
\begin{figure}[htbp]
	\centering
	\includegraphics[draft=false,width=.48\textwidth]{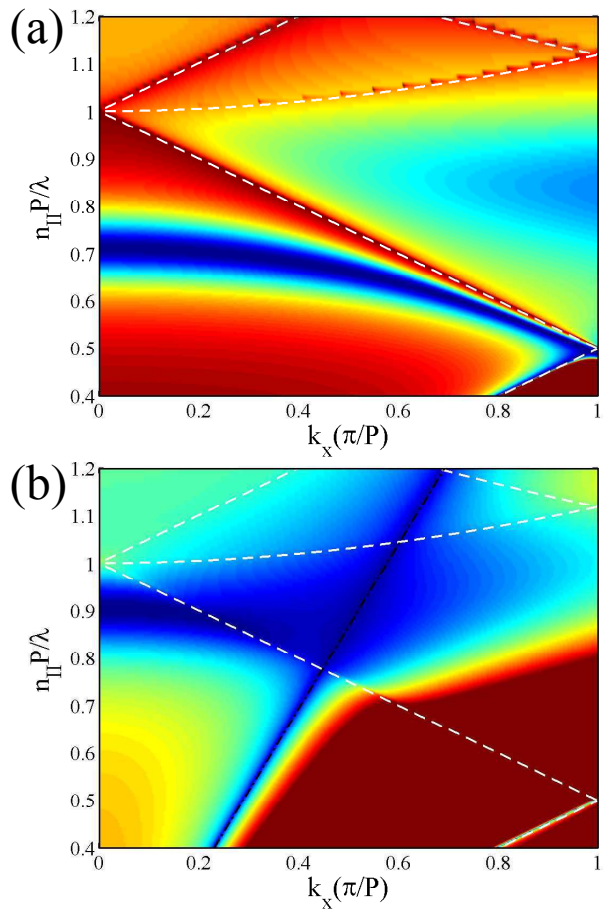}
	\caption{(Color online.) Angular-resolved zeroth-order (direct) transmission through the single-antenna metasurface in (a) symmetric ($\epsilon_{\rm I}=\epsilon_{\rm II}$) and (b) asymmetric ($\epsilon_{\rm I}=1, \epsilon_{\rm II}=12$) cladding. The parameters of the structure are the same as in Figs.~\ref{Fig:SAM} and~\ref{Fig:ComsolSAM}.}
	\label{Fig:DDAsym}
\end{figure}	

Figure~\ref{Fig:ComsolSAM} compares the results obtained with the use of the minimal model, the full MMT with a sufficient number of current modes (such that the convergence is reached), and the full-wave COMSOL Multiphysics simulations. The minimal model shows very good agreement with both of the exact techniques. The inclusion of the higher-order modes only insignificantly changes the spectra. Some discrepancy between the full MMT and COMSOL results is observed in the long wavelength range which is attributed to the rounding of the antenna corners that was done in COMSOL to minimize the numerical error due to spurious singularities at sharp corners. 

\subsection{Substrate effects on Wood's anomalies}
The substrate effect appears due to the finite dielectric contrast $\Delta\epsilon=\epsilon_\I-\epsilon_\II$ between the superstrate and substrate claddings, and in the MMT it is described by the wave admittances ($Y_{\vec{n},\tau}$) entering the expression for the sum $S_{11}$ [Eq.~(\ref{Eq:S})]. Predictions of the minimal model and the exact calculations clearly show that the presence of a substrate strongly affects the scattering characteristics of the antenna array (Fig.~\ref{Fig:SAM}). This effect can be fully understood from the analytical expressions of the minimal model. Let's first consider the case of a Wood's anomaly that indicates the onset of a diffraction order and appears in the transmission spectrum as a maximum (for symmetric cladding, $\epsilon_\I=\epsilon_\II$) or as a kink (in the general case) and represents the sharpest feature of the spectrum. Finite optical contrast between the substrate and the superstrate makes the sum $S_{11}$ converge at the onset of the $s$-polarized diffraction orders which manifests as the suppression of the Wood's anomalies. 

Indeed, as can be seen from the expression for $S_{11}(k_\|,\omega)$, in the case of a symmetric cladding ($\epsilon_\I$$=$$\epsilon_\II$) there is a divergent term corresponding to the onset of a $s$-polarized diffraction order $\ket{\vec{n},s}$; when the diffraction order experiences a transition between the evanescent and propagating regimes $Y_{\vec{n},s}\sim k_z^\vec{n}$ vanishes and the lattice sum $S_{11}(k_\|,\omega)$ diverges as $1/k_z^\vec{n}$. This gives rise to the Wood's anomaly and vanishing of the current mode in the antennas $c(k_\|,\omega)\rightarrow0$, as illustrated by black dashed curve in Fig.~\ref{Fig:SAM}a. Thus, at the Wood's anomaly, antennas are inactive and invisible ($t_{\vec{0},\tau}=1$ and $r_{\vec{0},\tau}=0$) as can be seen from Eqs.~(\ref{Eq:t1}-\ref{Eq:r1}). 

In the case of an asymmetric cladding, $\epsilon_\I\neq\epsilon_\II$, the same term of the lattice sum that was divergent for the symmetric case, assumes the form $1/(k_z^{\vec{n},{\rm I}}+k_z^{\vec{n},{\rm II}})$ and, since $k_z^{\vec{n},{\rm I}}\neq k_z^{\vec{n},{\rm II}}$, the lattice sum $S_{11}(k_\|,\omega)$ never diverges. As shown in Fig.~\ref{Fig:SAM}a by red dashed curve, even at the Wood's anomaly, the current in the wires $c(\kpar,\omega_{WA})$, does not vanish and the antenna array scatters the radiation. 
\begin{figure}[htbp]
	\centering
	\includegraphics[draft=false,width=.48\textwidth]{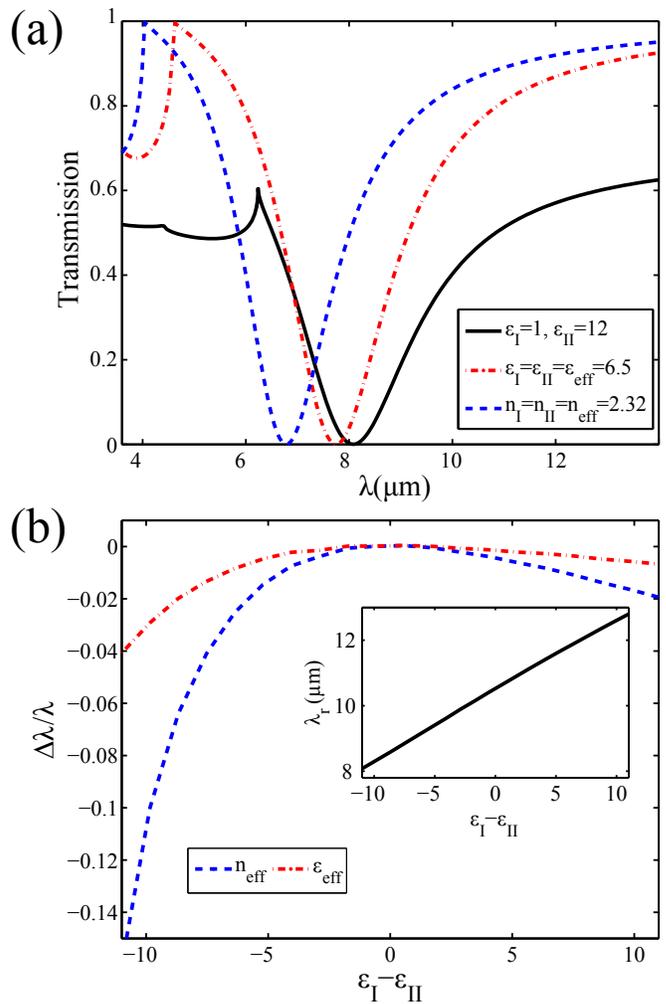}
	\caption{(Color online.) (a) Comparison between the exact treatment of the substrate and the results of the homogenized antenna's environment according to the effective medium theories: (i) $\epsilon_{eff}=(\epsilon_\I+\epsilon_\II)/2$ (dashed-dotted line) and (ii) $n_{eff}=(n_\I+n_\II)/2$ (dashed line). (b) Deviation of the resonance spectral position ($\Delta\lambda/\lambda_r$) predicted by the effective medium approaches from the exact result $\lambda_r$ shown in the inset.}
	\label{Fig:EffMedSAM}
\end{figure}		

The fact that antennas remain polarized and contribute to scattering at the Wood's anomaly for any finite value of the optical contrast $\Delta\epsilon$ is illustrated by Fig.~\ref{Fig:SAM}b, where the transmission at the wavelength corresponding to Wood's anomaly of the antenna array on the substrate (solid line) is plotted along with that of the bare (i.e., no antennas) interface (dashed line). The two curves intersect only for the case of a symmetric cladding and the discrepancy can be significant for a large contrast. Thus, the peak transmission of the antenna array on the substrate never reaches that of the bare interface. It is also remarkable that the transmission of the antenna array appears very asymmetric with respect to the sign of the contrast and the difference can reach tens of percent. This asymmetry is caused by the presence of the Wood's anomaly of the second medium whose diffraction order dominates in the regime of $\epsilon_\I>\epsilon_\II$.

\subsection{Substrate's influence on collective antenna resonances}
The spectral position of the collective dipolar antenna resonance corresponds to the minimum of the transmission or the maximum of the reflection spectra. It can also be found from the zeros of the imaginary part of the denominator in Eq.~\ref{Eq:c1}. In this paper, the length of the antennas are such that the observed collective mode stems from the $\lambda/2$ resonance of an individual antenna (as can be seen from the charge distribution plotted in the inset to Fig.~\ref{Fig:ComsolSAM}). As can be seen from dashed curves in Fig.~\ref{Fig:SAM}a, it also corresponds to the maximum of the current amplitude $c$ excited in the antennas. Note that a good agreement between the MMT and COMSOL spectra near the resonance in Fig.~\ref{Fig:ComsolSAM} indicates that our model fully accounts for the modified polarizability of the antennas caused by the presence of the substrate. 

The presence of the substrate and suppression of the Wood's anomalies dramatically affect the angular dispersion of the resonance and change its nature. Figure~\ref{Fig:DDAsym}(a) shows that for the case of a symmetric cladding we observe the effect of ``dragging'' of the mode by the Wood's anomaly\cite{garcia_de_abajo_colloquium:_2007,mousavi_suppression_2011}, which results in its strong spatial dispersion. This strongly dispersive character of the mode is a manifestation of its collective origin. Indeed, the long-range interaction among the individual metamolecules results in the appearance of the collective modes which are spectrally displaced and sharper as compared to those observed in the individual metamolecules~\cite{garcia_de_abajo_colloquium:_2007,mousavi_suppression_2011}. However, for the case of an asymmetric cladding, shown in the Fig.~\ref{Fig:DDAsym}b, the dragging effect is diminished and the resonance crosses the Wood's anomalies, every time reducing its quality factor due to the opening of additional radiative channels. The resulting non-dispersive behavior and lower quality factor of the mode imply that the resonance has lost its collective character and represent local excitation modified by the complex electromagnetic environment. Note that for the confined mode of the periodic metasurfaces, i.e. non-leaky spoof plasmons, presence of the substrate and associated suppression of the collective behavior may result in the disappearance of the mode~\cite{bendaa_confined_2009}.

By following the resonance one can also track its spectral position as a function of the dielectric contrast $\Delta \epsilon$. Inset to Fig.~\ref{Fig:EffMedSAM}b shows the corresponding spectral shift of the resonance. The frequency of the resonance shows nearly linear dependence on the contrast. We emphasize that the effect of the substrate on the antenna polarizability cannot be rigorously described by a basic dipole model, but can be phenomenologically incorporated into it. One of the most common approaches to include the optical contrast is to use an effective medium approach\cite{neubrech_resonances_2006} and approximate the permittivity or the refractive index of the background medium by $\epsilon_{eff}=(\epsilon_\I+\epsilon_\II)/2$ or $n_{eff}=(n_\I+n_\II)/2$. The validity of such approximations can be tested by comparing the spectral position of the resonance calculated for the antennas embedded into the effective medium with the exact MMT result. Figure~\ref{Fig:EffMedSAM}a shows the deviations of the resonance frequency as predicted by the effective medium theories from the exact result. One can see that the agreement between the effective medium approaches and the exact calculations is rather good. The homogenization of the permittivity $\epsilon_{eff}=(\epsilon_\I+\epsilon_\II)/2$ is closer to the exact result and provides a good estimate even for the case of a large contrast. However, it can be seen from Fig.~\ref{Fig:EffMedSAM}a that the effective medium approach fails in predicting the transmission and reflection, especially close to the Wood's anomaly.

\section{Fano resonance in double-antenna meta-surfaces}

	Next, the MMT for the double-antenna meta-surface (DAM) shown in Fig.~\ref{Fig:Schematics}b is considered. In this case, the truncation of the basis to the fundamental modes in both antennas, results in a minimal model with a 2x2 matrix equation for the amplitudes $c_1$ and $c_2$:
\begin{align}
\begin{bmatrix}
 S_{11} & S_{12}\\
 S_{21} & S_{11}
\end{bmatrix}
\begin{bmatrix}
  c_1\\
  c_2
\end{bmatrix}
=\chi E_{\rm ext}\begin{bmatrix}
  e^{-ik_xD_x/2}\\
  e^{+ik_xD_x/2}
\end{bmatrix},
\end{align}	
where $S_{mm\prime}\equiv S_{11}^{mm\prime}$ represents the effective Green's function given by Eq.~(\ref{Eq:S}), and $\chi$=$2 \YI{0}/(\YI{0}+\YII{0} )$ $\exp(i k_x D_x/2)\braket{\vec{j}^{(1)}}{\vec{0},\tau}$ is the coupling strength of the incident light ($\ket{\vec{0},\tau}$) to the fundamental current modes.

\begin{figure}[htbp]
	\centering
	\includegraphics[draft=false,width=.48\textwidth]{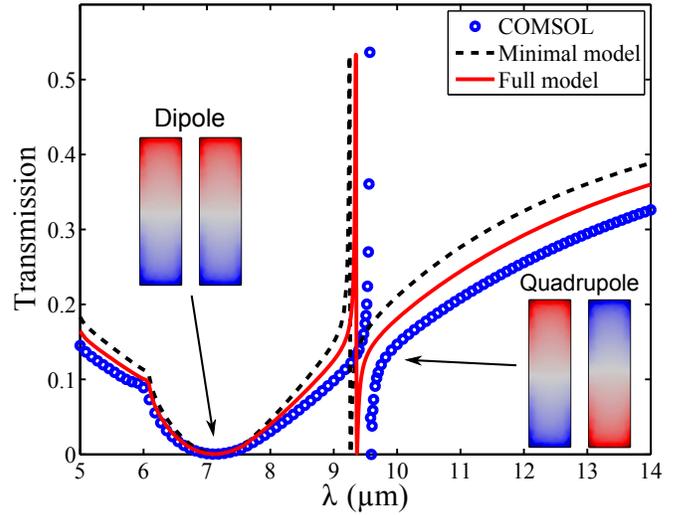}
	\caption{(Color online.) Comparison of the zeroth-order transmission spectra of a double-antenna metasurface on the substrate for 50$^\circ$ angle of $s$-polarized incidence calculated with the use of COMSOL Multiphysics, and the minimal and full MMT models. The structure parameters are as follows: $a_x=0.3~{\mu}m$, $a_y=1.5~{\mu}m$, $P_x=P_y=1.8~{\mu}m$, $D_x=0.6~{\mu}m$, $\epsilon_\I=1$, and $\epsilon_\II=12$.}
	\label{Fig:ComsolDAM}
\end{figure}		

Using the unitary transformation
\begin{align}
\csub&=c_1\,\exp(ik_x D_x/2)-c_2\,\exp(-ik_x D_x/2),\notag\\
\csup&=c_1\,\exp(ik_x D_x/2)+c_2\,\exp(-ik_x D_x/2),
\end{align}
the basis of the two fundamental electric currents can be changed to the more instructive basis of the sub-radiant and super-radiant current modes. In the new basis, the system is described by the matrix equation 
\begin{align}
\begin{bmatrix}
  S_{11}+\Delta & i\kappa\\
  -i\kappa & S_{11}-\Delta
\end{bmatrix}
\begin{bmatrix}
  \csup\\
  \csub
\end{bmatrix}
=\chi E_{\rm ext}\begin{bmatrix}
  2\\
  0
\end{bmatrix},
\label{Eq:SubSup}
\end{align}
where $\Delta$=$-1/2[S_{12} \exp(ik_x D_x)+S_{21} \exp(-ik_x D_x)]$ and $\kappa$=$i/2[S_{12}  \exp(ik_x D_x)-S_{21} \exp(-ik_x D_x)]$. These modes are the result of the radiative coupling of the dipolar resonances of the antenna pairs~\cite{liu_plasmonic_2009,wu_broadband_2011,mousavi_suppression_2011}. The super-radiant and sub-radiant modes have distinct symmetric and anti-symmetric charge distributions, while the currents corresponding to them are collinear and anti-collinear, respectively, as illustrated by the inset to Fig.~\ref{Fig:ComsolDAM}. The modes interaction makes them blue-shifted (super-radiant) and red-shifted (sub-radiant) with respect to the electric-dipolar resonance of SAM. Another consequence of such hybridization and different symmetry of the modes is their different radiative coupling efficiency. The sub-radiant mode is not directly coupled to the incident light. The latter fact is reflected in the mode's name and directly follows from the zero radiative coupling strength on the RHS of Eq.~(\ref{Eq:SubSup}).
\begin{figure}[htbp]
	\centering
	\includegraphics[draft=false,width=.48\textwidth]{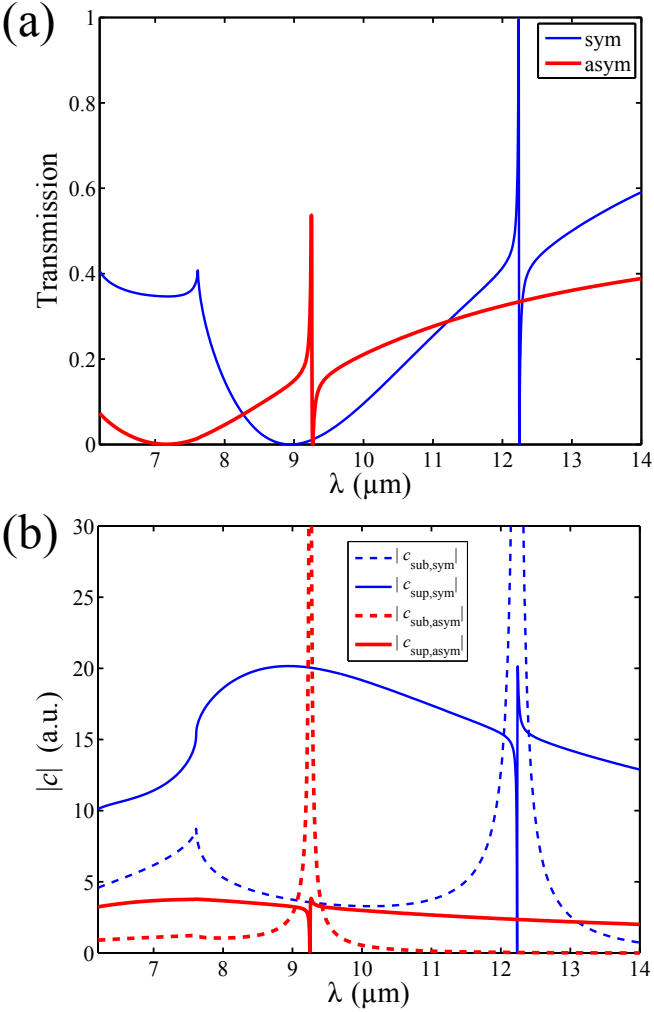}
	\caption{(Color online.) (a) Zeroth-order $s$-polarized transmission spectra of the double-antenna meta-surface for the symmetric ($\epsilon_{\rm I}$=$\epsilon_{\rm II}$=12, blue line) and asymmetric ($\epsilon_{\rm I}$=1, $\epsilon_{\rm II}$=12, red line) claddings. (b) Amplitude of the sub-radiant (dashed lines) and super-radiant (solid lines) current modes. The parameters of the structure and the incidence angle are the same as in Fig.~\ref{Fig:ComsolDAM}.}
	\label{Fig:DAM}
\end{figure}	
\begin{figure}[htbp]
	\centering
	\includegraphics[draft=false,width=.48\textwidth]{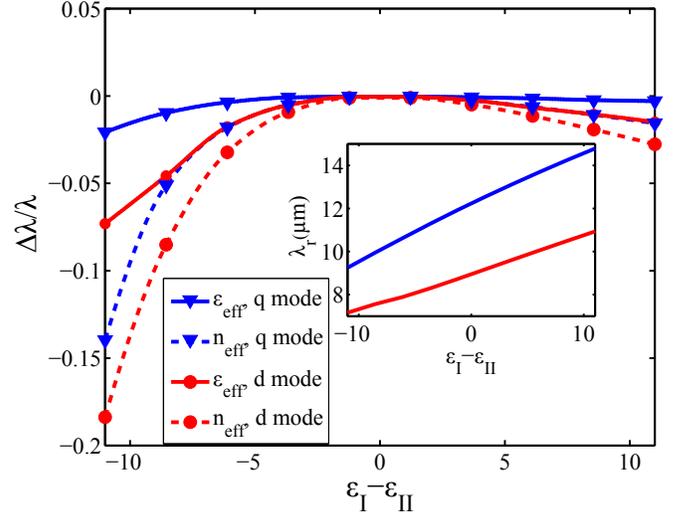}
	\caption{(Color online.) Deviation of the spectral position ($\Delta\lambda/\lambda_r$) of the dipolar (red lines with circle markers) and quadrupolar (blue lines with triangle markers) resonances predicted by the effective medium approaches from the exact results $\lambda_r$ shown in the inset.}
	\label{Fig:EffMedDAM}
\end{figure}

 With the use of Eqs.~(\ref{Eq:tn},\ref{Eq:rn}) we obtain a set of minimal-model expressions for the transmission and reflection coefficients:
 \begin{align}
		t_{\vec{n},\tau}=\frac{2\YI{n}}{\YI{n}+\YII{n}}i_{\vec{n},\tau}-\frac{\braket{\vec{n},\tau}{\vec{j}^{(1)}}\exp{(-ik_xD_x/2)}}{\YI{n}+\YII{n}}\csup,\label{Eq:t2}\\
		r_{\vec{n},\tau}=\frac{\YI{n}-\YII{n}}{\YI{n}+\YII{n}}i_{\vec{n},\tau}-\frac{\braket{\vec{n},\tau}{\vec{j}^{(1)}}\exp{(-ik_xD_x/2)}}{\YI{n}+\YII{n}}\csup.\label{Eq:r2}
	\end{align}

Note that because of their coupling, $\csup$ and $\csub$ are not the true eigenmodes of the system. The true eigenmodes which will be referred to as dipolar $d$ and quadrupolar $q$ can be obtained using the transformation:\cite{mousavi_suppression_2011}
\begin{align}
d=\frac{\Delta+X}{2X}\csup\,+\,\frac{i\kappa}{2X}\csub,\notag\\
q=\frac{i\kappa}{2X}\csup\,+\,\frac{\Delta+X}{2X}\csub, \label{Eq:dq1}
\end{align}
where $X=\sqrt{\Delta^2+\kappa^2}$. The modal coefficients $d$ and $q$ are related to the incident field through 
\begin{align}
d=&\frac{1+\Delta/X} {S_d} \chi E_{ext},\quad S_d=S_{11}+X, \notag\\
q=&\frac{i\kappa/X}{S_q}\chi E_{ext},\quad S_q=S_{11}-X, \label{Eq:dq2}
\end{align}

In the particular case of normal incidence, the quadrupolar mode $q$ exactly coincides with the sub-radiant mode $\csub$ since its coupling to the super-radiant mode $\csup$ vanishes ($\kappa\rightarrow0$ as $\kpar\rightarrow0$). However, at finite angles of incidence, this mode acquires a finite electric-dipolar moment and its radiative coupling and bandwidth gradually increase. The dipolar mode, in contrast, is always strongly radiatively coupled and is spectrally broad at any incidence angle. Eqs.~(\ref{Eq:t2}-\ref{Eq:r2}) describe a Fano resonance of a single "continuum" interacting with two "discrete" dipolar ($d$) and quadrupolar ($q$) resonances. However, the bandwidth of the last two is very different due to their different radiative coupling. 

%Zeroth-order transmission spectrum calculated with the use of minimal model for both cases of symmetric claddings and asymmetric claddings is plotted in Fig.~\ref{Fig:DAM}

The transmission spectrum of DAM for an oblique incidence of 50$^{\circ}$ calculated using the minimal model, full MMT and COMSOL solver are plotted alongside in Fig.~\ref{Fig:ComsolDAM}. As before, the different techniques show very good agreement, which implies the applicability of the analytical minimal model even for the description of light scattering by more complex metasurfaces such as DAM. Two resonances, a broad one due to the excitation of the dipolar mode and a narrow one due to the excitation of the quadrupolar mode, are clearly seen in the spectra. As expected, the narrow resonance has a strongly asymmetric shape typical for Fano resonances. 

\subsection{Dipolar mode}
Because the dipolar and quadrupolar resonances are spectrally separated and the quadrupolar resonance is spectrally narrow, first we can consider the dipolar resonance individually. It has been recently shown that the collective dipolar resonance in DAM appears to have spatial dispersion [$\omega(\vec{k}_\|)$] very different from that found in SAM ~\cite{mousavi_suppression_2011}. 
Using the basic dipole model, this difference was shown to be due to the suppression of the long-range interactions in DAMs. MMT provides analogous results. In the present case, however, the role of the substrate and the antenna geometry are fully taken into account. In DAM, the Wood's anomalies would have corresponded to the divergences (or local maxima for an asymmetric cladding) of the effective Green's function $S_d=S_{11}+X$ playing the role of the lattice sum in the dipole model or the Green's function $S_{11}$ of SAM. However, it can be shown that exactly at the frequency where the Wood's anomaly is expected, the diverging terms in $S_{11}$ and $X$ exactly cancel out eliminating the divergence, thereby suppressing the spectral features associated with the Wood's anomaly~\cite{mousavi_suppression_2011}. Thus, in the DAM case, Wood's anomaly appears to be suppressed even for a symmetric cladding ($\Delta\epsilon=0$). Therefore for the asymmetric cladding, the Wood's anomaly is suppressed by two mechanisms: one due to substrate/superstrate contrast just as in the case of SAM, and another one due to disconnected topology of DAM. Note, however, that despite such suppression, one might still observe some variations of reflectivity in the vicinity of the expected Wood's anomalies, which is a result of the opening of the diffraction channel and change of the mode's radiative lifetime.~\cite{mousavi_suppression_2011}.

\subsection{Quadrupolar mode}
Now we focus on the quadrupolar resonance. The narrow spectral region where the transmission undergoes a rapid variation is especially interesting because of the strong field enhancement and slow light regime reported earlier~\cite{fedotov_sharp_2007, wu_broadband_2011}. The asymmetric shape of the spectrum is the result of the Fano interference between the quadrupolar resonance and a background provided by two transmission channels: (i) the direct transmission without interaction with the array, and (ii) the radiation scattered by the dipolar resonance. This background is nearly constant and is featureless over the spectral width of the quadrupolar resonance where the transmission changes from its maximal value to zero. It can be analytically shown that the transmission peak corresponds to the divergence of $\csub$ and zero of $\csup$(Fig.~\ref{Fig:DAM}b). Because only the super-radiant mode ($\csup$) is coupled to the radiation, the antennas do not radiate at this frequency and the transmission acquires a universal value of the transmission of the bare interface between the substrate and the superstrate as seen from Eqs.~(\ref{Eq:t2},\ref{Eq:r2}). Note that this behavior is different from that found for another type of Fano resonances, Wood's anomalies in SAM, when for an asymmetric cladding configuration, the radiative current mode always had some finite amplitude. %This fact is a result of different nature of the Fano resonances. In the present case the Fano resonance
	
	In the previous section, it was demonstrated that the effective medium approach accurately predicted the spectral position of the dipolar resonance for SAM. Here we test this approach for both cases of the dipolar and quadrupolar resonances in DAM. At first, the inset to Fig.~\ref{Fig:EffMedDAM} shows the spectral position of the resonances calculated by the full MMT approach as a function of the dielectric contrast, and Fig.~\ref{Fig:EffMedDAM} shows the deviation from the exact result. The effective theories match well with the exact results for both quadrupolar and dipolar resonances, while the permittivity homogenization $\epsilon_{\rm eff}=(\epsilon_\I+\epsilon_\II)/2$ is more accurate than the index homogenization $n_{\rm eff}=(n_\I+n_\II)/2$. However, just as in the case of SAM, none of the homogenization approaches succeed in predicting the transmission/reflection spectra.
	
	Finally, we observe that in contrast to the dipolar mode, the quadrupolar mode in DAM is strongly affected by the Wood's anomaly. From the expression for its amplitude Eq.~(\ref{Eq:dq2}) one can see that for the quadrupolar mode the divergence in the Green's function $S_{11}-X$ does take place in the case of a symmetric cladding $\epsilon_\I=\epsilon_\II$. The effect of the Wood's anomaly on the quadrupolar resonance of the DAM is expected to be especially significant for large angles of incidence, when they approach each other. This results in a strong spatial dispersion of the quadrupolar mode observed earlier both in mid- and near-IR plasmonic DAMs ~\cite{mousavi_suppression_2011}. For the case of an asymmetric cladding the effect of the Wood's anomaly on the quadrupolar resonance is again suppressed due to the mismatch of the wave admittances in the substrate and the superstrate and all the arguments about the suppression of the Wood's anomalies used in the previous section for SAM are applicable here.
\citeindexfalse
\section{Electromagnetically Induced Transparency}
In this section a periodic dolmen structure with three antennas in its unit cell~\cite{zhang_plasmon-induced_2008}$^,$\cite{papasimakis_metamaterial_2008}$^,$\cite{liu_plasmonic_2009}$^,$\cite{verellen_fano_2009}$^,$\cite{wu_broadband_2011} [shown in Fig.~\ref{Fig:Schematics}c] is studied. This design was first introduced to mimic EIT in plasmonic structures for normally incident $x$-polarized light~\cite{zhang_plasmon-induced_2008}. Two vertical antennas on the dolmen structure are responsible for formation of the quadrupolar mode analogous to that described in the previous section. This mode plays the role of the sub-radiant mode of EIT. The horizontal antenna provides a spectrally broad dipolar response. Its length is chosen in such a way that the frequency of the dipolar resonance is matched to the frequency of the sub-radiant (quadrupolar) mode. The position of the horizontal antenna within the unit cell defines the degree of symmetry breaking of the metamolecule and the intensity of coupling between the two modes. The sub-radiant mode is completely decoupled from $x$-polarized incident radiation if the antenna is placed equidistantly between the two nearest vertical double antennas, $D_y=0$. However, as soon as the horizontal antenna is displaced vertically from this position ($D_y\neq0$), the sub-radiant mode couples to the incident radiation indirectly through the dipolar mode of the horizontal antenna. This coupling to the sub-radiant mode results in a transmission/reflection spectrum typical for the EIT: a sharp transmission resonance embedded into a region of highly reflecting background provided by the dipolar antenna mode. Note that for the polarization under study the dipolar mode of two-vertical antennas is not excited since there is no $y$-component of the incident electric field. In principle, this mode could be excited indirectly provided that the symmetry of the metamolecule is reduced further by displacing the horizontal antenna in the x-direction~\cite{wu_broadband_2011}. However, this situation will not be considered here.

Within the minimal model, limiting the current basis to only fundamental current mode for each antenna and neglecting finite conductivity of the metallic antennas ($1/\sigma=0$), the EIT structure can be described by a 3x3 matrix equation relating the current amplitudes to the incident electric field:
	\begin{equation}
		\begin{bmatrix}
		   S_{11}  &  S_{12} &  S_{13}\\
		   S_{12}  &  S_{11} & -S_{13}\\
		   S_{13}   & -S_{13} & S_{33}
	\end{bmatrix} 
	\begin{bmatrix}
		c_1 \\ c_2 \\ c_3
	\end{bmatrix}= \chi E_{\rm ext}\begin{bmatrix}
		0 \\ 0 \\ 1
	\end{bmatrix},
	\label{Eq:M33}
	\end{equation}
where $S_{m m\prime}\equiv S_{11}^{m m\prime}$ represents the effective Green's function given by Eq.~(\ref{Eq:S}), and $\chi=<j_3|0>2\YI{0}/(\YI{0}+\YII{0})$ is the coupling strength between the incident light and the horizontal antenna (labeled as ``3''). The RHS of Eq.~(\ref{Eq:M33}) reflects the fact that the vertical antennas are not coupled to the $x$-polarized incident light. On the other hand, the presence of off-diagonal matrix elements in the LHS of this equation indicates that they are coupled to the horizontal antenna. It follows from the symmetry of the structure that the basis of three currents is redundant for the case of $x$-polarized normal incidence, under which the dipolar mode of the antenna pair $D_y$ cannot be excited. By performing the unitary transformation $D_y=c_1+c_2$, $Q=c_1-c_2$ and $D_x=c_3$, the non-interacting $D_y$ mode can be eliminated. The truncated governing equation for the super-radiant mode $D_x$ and the sub-radiant mode $Q$ assumes the form: 
\begin{figure}[htbp]
	\centering
	\includegraphics[draft=false,width=.48\textwidth]{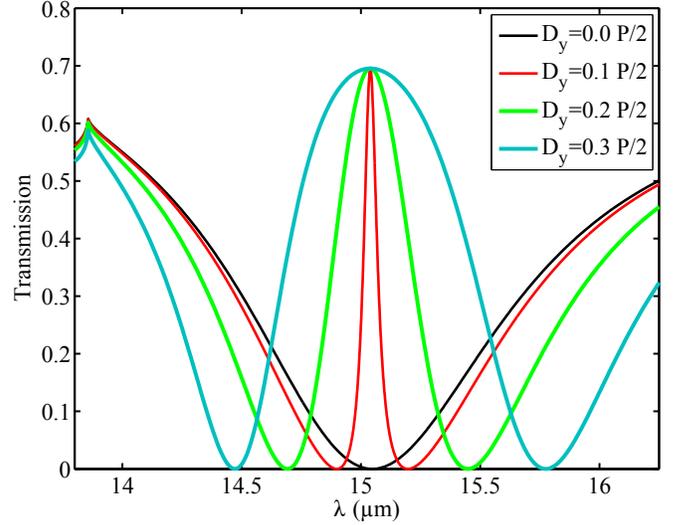}
	\caption{(Color online.) Substrate effect on the normal-incidence zeroth-order transmission spectra of the dolmen metasurface for different values of symmetry-breaking parameter $D_y$. Horizontal antenna's dimensions are $a_x=1.79\,\mu$m, $a_y=0.6\,\mu$m. Vertical antennas are identical with $a_x=0.6\,\mu$m and $a_y=1.7\,\mu$m and separated by $D_x=1.2\,\mu$m. $\epsilon_\I=1$, $\epsilon_\II=12$, and $P_x$$=$$P_y$$=$$4\,\mu$m.}
	\label{Fig:EIT}
\end{figure}
\begin{equation}
\begin{bmatrix}
   S_Q  &  \kappa \\
   \kappa & S_{D_x}
\end{bmatrix} 
\begin{bmatrix}
Q \\ D_x
\end{bmatrix}= \chi E_{\rm ext}\begin{bmatrix}
0 \\ 1
\end{bmatrix},
\label{Eq:M22}
\end{equation}
where $S_Q=1/2(S_{11}-S_{12})$, $S_{D_x}=S_{33}$, and the coupling between the modes is given by $\kappa=S_{13}$. 
Note, that unless the horizontal antenna is placed symmetrically (in which case $\kappa=0$), the modes $D_x$ and $Q$ are not eigenmodes of the EIT structure, since they are coupled to each other through the off-diagonal matrix elements in the LHS of Eq.~(\ref{Eq:M22}). This coupling gives rise to hybridization of the sub-radiant mode $Q$ and the super-radiant mode $D_x$ and formation of new mixed states -- quasi-dipolar $\tilde{D}_x$ and quasi-quadrupolar $\tilde{Q}$. Both of these modes are coupled to the incident electromagnetic field, but have very disparate coupling efficiency. The true eigenmodes of the structure can be found through an eigen-decomposition procedure, $\tilde{D}_x=1/(1+\eta^2)\,D_x+\eta/(1+\eta^2)\,Q$ and $\tilde{Q}=1/(1+\eta^2)\,Q-\eta/(1+\eta^2)\,D_x$ where $\eta=(S_Q-S_{D_x}-\Delta)/2\kappa$ [that tends to zero in the case of a vanishing coupling $\kappa\rightarrow0$] and $\Delta=\sqrt{(S_Q-S_{D_x})^2+4\kappa^2}$. 
Then the equations for the eigenmodes' amplitudes assume the form
\begin{eqnarray}
\tilde{D}_x&=& \frac{2/(1+\eta^2)}{S_Q+S_{D_x}-\Delta}\,\chi E_{\rm ext}, \\
\tilde{Q}&=& \frac{-2\eta/(1+\eta^2)}{S_Q+S_{D_x}+\Delta}\,\chi E_{\rm ext},
 \label{Eq:DxQ}
\end{eqnarray}
and the reflection/transmission coefficients of the structure can be found from the following equations:
\begin{align}
	r_{\vec{n},\tau}&=\frac{\YI{n}-\YII{n}}{\YI{n}+\YII{n}}+\frac{\braket{\vec{n},\tau}{\vec{j}_3}}{\YI{n}+\YII{n}}\;D_x,\\
	t_{\vec{n},\tau}&=\frac{2\YI{n}}{\YI{n}+\YII{n}}+\frac{\braket{\vec{n},\tau}{\vec{j}_3}}{\YI{n}+\YII{n}}\;D_x,\\
	D_x&=\tilde{D}_x-\eta\,\tilde{Q}.
\end{align}

The transmission spectra of the structure are shown in Fig.~\ref{Fig:EIT}. The different curves demonstrate how the radiative coupling of the sub-radiant mode increases when the symmetry of the metamolecule is gradually reduced by increasing the parameter $D_y$. This manifests, at first, as an appearance of the sharp and narrow EIT transmission peak and a gradual increase of the EIT peak bandwidth. It is remarkable that regardless of the symmetry breaking degree, the maximum of the EIT peak always tends to the same value defined by the transmission of the bare dielectric interface between the substrate and superstrate. Therefore we can claim that this is rather general rule for all Fano resonant systems, where a Fano resonance originates from the structure of the metamolecule and not from Wood's anomaly as in the case of SAM.

\section{Conclusions}
We investigated the effects of a substrate on three different mid-IR Fano metasurfaces and found that a finite refraction-index contrast between the top and bottom claddings dramatically changes the optical response of the structures. In addition to the expected spectral shift of the resonances induced by the substrate, a dramatic change in the collective behavior of the systems was found. The widely used effective medium techniques were tested. While satisfactorily predicting the spectral positions of the resonances, they failed to describe other scattering characteristics of the structures. The Wood's anomalies and modes' dispersion were found to be strongly affected by the presence of a substrate. Divergences corresponding to the onset of the $s$-polarized diffraction orders disappear whenever the refractive indices of the substrate and the superstrate are mismatched, resulting in the suppression of the spectral features associated with the Wood's anomalies and giving rise to anomalously flat dispersion of the modes. As the Wood's anomaly and the surface resonances of the periodic metasurfaces can be considered as a Fano resonance due to the collective interaction of metamolecules, it can be concluded that the presence of a substrate destroys the Fano picture. As a consequence, the dramatic effect of the onset of the diffraction orders reduces to the less dramatic change of the modes radiative lifetime. The nature of the modes also changes as they lose their collective character and start resembling local resonances yet modified by their complex environment. 

In addition to this collective Fano resonance, we also studied the effect of the substrate on Fano resonances originating from the local geometry of the metamolecules. Two types of Fano systems were considered with a sharp quadrupolar resonance detuned from (double-antenna metasurface) and tuned to (dolmen metasurface) the dipolar resonance providing the Fano background. In both systems the Fano resonances survived after the introduction of the substrate but were affected due to the modification of the frequencies of both low-$Q$ (dipolar) and high-$Q$ (quadrupolar) resonances as well as additional reflection from the substrate. A universal behavior was discovered for both systems: (a) the peak transmission reached the value corresponding to the bare interface, and (b) antennas became effectively invisible at the frequency of the Fano resonance. These findings will be useful for designing non-dispersive photonic devices with a wide-angle optical response.

\acknowledgments
This research was supported by the Office of Naval Research (grant N00014-10-1-0929), the Air Force Office of Scientific Research (grant FA8650-090-D-5037), and the National Science Foundation (grant CMMI-0928664).
\bibliography{OurBib}

\end{document}